\documentclass[a4paper,10pt]{article}
\usepackage[left=2.35cm,right=2.35cm,top=2.54cm,
headheight=0.5cm,headsep=0.54cm,bottom=2.54cm,footskip=0.79cm
]{geometry}

\usepackage{amsmath,amssymb,dsfont}

\usepackage{hyperref}

\usepackage{cite}

\begin{document}

\title{Connes spectral distance and nonlocality of generalized noncommutative phase spaces}

\author{Bing-Sheng Lin$^{1,2,\dag}$,\,\,Tai-Hua Heng$^{\,3}$\\
	\small $^{1}$School of Mathematics, South China University of Technology, 
	\small Guangzhou 510641, China\\
	\small$^{2}$Laboratory of Quantum Science and Engineering,\\
	\small South China University of Technology, Guangzhou 510641, China\\
	\small $^{\dag}$Email: sclbs@scut.edu.cn\\
	\small $^{3}$School of Physics and Optoelectronic Engineering, Anhui University, Hefei 230601, China}

\date{}

\maketitle

\begin{abstract}
We study the Connes spectral distance of quantum states and analyse the nonlocality of a 4D generalized noncommutative phase space. By virtue of the Hilbert-Schmidt operatorial formulation, we obtain the Dirac operator and construct a spectral triple corresponding to the noncommutative phase space. Based on the ball condition, we obtain some constraint relations about the optimal elements, and then calculate the Connes spectral distance between two Fock states. Due to the noncommutativity, the spectral distances between Fock states in generalized noncommutative phase spaces are shorter than those in normal phase spaces. This shortening of distances implies some type of nonlocality caused by the noncommutativity. These spectral distances in the 4D generalized noncommutative phase space are additive and satisfy the normal Pythagoras theorem. When the noncommutative parameters go to zero, the results return to those in normal quantum phase spaces.
\end{abstract}

\section{Introduction}
The ideas of noncommutative spacetime started in 1947 \cite{Snyder}. In the 1980s,
Connes formulated the mathematically rigorous framework of noncommutative
geometry \cite{Connes}.
In the past decades, there has been much interest in the study of physics in noncommutative
spaces \cite{Seiberg,Douglas,Gayral,Chaichian,Ettefaghi,lhc,Calmet,Couch,Muhuri}.
A noncommutative spacetime also appeared in string theory, namely in the quantization of open string \cite{Seiberg}. The noncommutativity
of spacetime also plays an important role in quantum gravity \cite{Doplicher,Zupnik}. The concept of noncommutative spacetime is also applied in condensed matter physics, such as the integer quantum Hall effect \cite{Polychronakos}. Usually, only spatial noncommutativity is considered.
But many researchers have also studied models in which a noncommutative
geometry is defined on the whole phase space \cite{Duval,Nair,Banerjee,Zhang,Li,Bastos,HMS,Jing,Lin,Gnatenko}. Noncommutativity between momenta can be naturally considered as a consequence of noncommutativity
between positions, as momenta are defined to be the partial derivatives of the action with respect to the position coordinates.

Because of the noncommutativity, there are no traditional point and distance in a noncommutative space. Instead one can study states in noncommutative spaces. In a noncommutative space, a pure state is the analog of a traditional point in a normal commutative space. One can calculate some kinds of distance measures between the states, such as the Connes spectral distance \cite{Connes1}.
The Connes spectral distances in some kinds of noncommutative spaces have already been studied in the studies
\cite{Bimonte,Cagnache,Martinetti,DAndrea0,DAndrea,Franco,Scholtz,Chaoba,
Revisiting,Kumar,Chakraborty,Barrett}.
For example, Cagnache \emph{et. al.} have studied the Connes spectral distance in the Moyal plane \cite{Cagnache}. They explicitly computed Connes spectral distance between the pure states which corresponding to eigenfunctions of the quantum harmonic oscillators.
Scholtz and his collaborators have developed the Hilbert-Schmidt operatorial formulation \cite{Scholtz,Chaoba,Revisiting}, they studied the Connes spectral distances of harmonic oscillator states and also coherent states in Moyal plane and fuzzy space.
Barrett \emph{et al.} also used Monte Carlo simulation to study spectral distances in the fuzzy spaces \cite{Barrett}.

In the present work, we study the Connes spectral distance of the quantum states in a 4D generalized noncommutative phase space (NCPS). We find that the Connes spectral distances of the quantum states in the noncommutative phase space are shorter than those in a normal phase space. This means that the noncommutativity of the generalized noncommutative phase space can lead to some type of nonlocality. This nonlocality can equivalently make the distances shorter.

This paper is organized as follows. In Sect.~\ref{sec2}, we construct a spectral triple corresponding to a 4D generalized noncommutative phase space. Using the Hilbert-Schmidt operatorial formulation, we construct a boson Fock space and a quantum Hilbert space, and obtain the Dirac operator.
In Sect.~\ref{sec3}, we review the definition of Connes spectral distance. Based on the ball condition, we derive some constraint relations about the optimal elements.
The Connes spectral distances between Fock states in the generalized noncommutative phase space are calculated in Sect.~\ref{sec4}.
Some discussions and conclusions are given in Sect.~\ref{sec5}. Some details of calculations of the Dirac operator and optimal elements are given in Appendixes.

\section{Spectral triple of generalized noncommutative phase space}\label{sec2}
Now let us consider the following 4D generalized noncommutative phase space \cite{Lin1}, the position operators $\hat{X}_i$ and the momentum operators $\hat{P}_i$ satisfy the following commutation relations,
\begin{equation}
[\hat{X}_1,\,\hat{P}_1]=[\hat{X}_2,\,\hat{P}_2]=\mathrm{i}\hbar\,,\qquad
[\hat{X}_1,\,\hat{X}_2]=\mathrm{i}\mu\,,\qquad
[\hat{P}_1,\,\hat{P}_2]=\mathrm{i}\nu\,,
\end{equation}
and others vanish. Here $\mu$, $\nu$ are some real parameters. In the present work, we only consider the case $\mu, \nu > 0$.
For simplicity, we can use the following transformations \cite{squeezed},
\begin{equation}
\tilde{X}_i=\sqrt[4]{\frac{\nu}{\mu}}\hat{X}_i,\qquad
\tilde{P}_i=\sqrt[4]{\frac{\mu}{\nu}}\hat{P}_i,
\end{equation}
and we have
\begin{equation}\label{ncps}
[\tilde{X}_1,\,\tilde{P}_1]=[\tilde{X}_2,\,\tilde{P}_2]=\mathrm{i}\hbar\,,\qquad
[\tilde{X}_1,\,\tilde{X}_2]=\mathrm{i}\theta\,,\qquad
[\tilde{P}_1,\,\tilde{P}_2]=\mathrm{i}\theta\,,
\end{equation}
and others vanish. Here $\theta=\sqrt{\mu\nu}$. Usually we also assume $\theta\ll\hbar$.

In a normal 4D quantum phase space,  the coordinate operators $\hat{x}_i$, $\hat{p}_i$ satisfy the following commutation relations,
\begin{equation}\label{qps}
[\hat{x}_i,\,\hat{p}_j]=\mathrm{i}\delta_{ij}\hbar,\qquad i,j=1,2,
\end{equation}
and others vanish.
Similarly, we can denote
\begin{equation}
\tilde{x}_i=\sqrt[4]{\frac{\nu}{\mu}}\,\hat{x}_i,\qquad
\tilde{p}_i=\sqrt[4]{\frac{\mu}{\nu}}\,\hat{p}_i,
\end{equation}
and there are the similar commutation relations $[\tilde{x}_i,\,\tilde{p}_j]=\mathrm{i}\delta_{ij}\hbar$. Define the creation and annihilation operators
\begin{equation}
\hat{a}_i=\frac{1}{\sqrt{2\hbar}}(\tilde{x}_i
+\mathrm{i}\tilde{p}_i),\qquad
\hat{a}_i^{\dag}=\frac{1}{\sqrt{2\hbar}}(\tilde{x}_i
-\mathrm{i}\tilde{p}_i),
\end{equation}
these operators satisfy the commutation relations $[\hat{a}_i,\hat{a}_j^{\dag}]=\delta_{ij}$, $[\hat{a}_i,\hat{a}_j]=[\hat{a}_i^{\dag},\hat{a}_j^{\dag}]=0$.
One can define the Fock states as follows,
\begin{equation}
|n\rangle_{\!i}
=\frac{1}{\sqrt{n!}}(\hat{a}_i^{\dag})^{n}|0\rangle,
\qquad i=1,2,\quad n=0,1,2,...\,,
\end{equation}
where $|0\rangle$ is the vacuum state, and $\hat{a}_i|0\rangle=0$.
These Fock states $|n\rangle_{\!i}$ satisfy the relations
\begin{eqnarray}\label{a1a2}
&&\hat{a}_i|n\rangle_{\!i}=\sqrt{n}|n-1\rangle_{\!i},\qquad \hat{a}_i^{\dag}|n\rangle_{\!i}=\sqrt{n+1}|n+1\rangle_{\!i}.
\end{eqnarray}
We also have the resolution of the identity,
\begin{eqnarray}
&&\sum_{n=0}^{\infty}|n\rangle_{\!i\,i\!}\langle n|=\mathds{I}_i.
\end{eqnarray}

Furthermore, one can define a boson Fock space as follows,
\begin{equation}
\mathcal{F}=\mathrm{span}\left\{|m,n\rangle
:=|m\rangle_{\!1}\otimes|n\rangle_{\!2},
\quad m,n=0,1,2,...\right\}.
\end{equation}
There are the relations,
\begin{eqnarray}
&&\sum_{n=0}^{\infty}|m,n\rangle\langle m,n|=|m\rangle_{\!1\,1\!}\langle m|\otimes\mathds{I}_2,\qquad
\sum_{m=0}^{\infty}|m,n\rangle\langle m,n|=\mathds{I}_1\otimes|n\rangle_{\!2\,2\!}\langle n|,\nonumber\\
&&\sum_{m,n=0}^{\infty}|m,n\rangle\langle m,n|=\mathds{I}_1\otimes\mathds{I}_2.
\end{eqnarray}
By virtue of the Hilbert-Schmidt operatorial formulation \cite{Formulation}, one can construct a quantum Hilbert space as follows,
\begin{equation}
\mathcal{Q}=\mathrm{span}\left\{|m_1,m_2\rangle\langle n_1,n_2|\right\}.
\end{equation}
The elements $\psi$ of the quantum Hilbert space $\mathcal{Q}$ are denoted by $|\psi)$,
and the inner product is defined as
\begin{equation}
(\phi|\psi)
:=\mathrm{tr}_\mathcal{F}\left(\phi^{\dag}
\psi\right)
=\sum_{m,n=0}^{\infty}\langle m,n|\phi^{\dag}\psi|m,n\rangle,
\end{equation}
where $\mathrm{tr}_{\mathcal{F}}(\cdot)$ denotes the trace over $\mathcal{F}$.

In general, a noncommutative space corresponds to a spectral triple $(\mathcal{A},\mathcal{H},\mathcal{D})$ \cite{Connes}, where $\mathcal{A}$ is an involutive algebra acting on a Hilbert space $\mathcal{H}$, and $\mathcal{D}$ is the Dirac operator on $\mathcal{H}$.
One can construct a spectral triple $(\mathcal{A},\mathcal{H},\mathcal{D})$ corresponding to the generalized noncommutative phase space (\ref{ncps}) as follows,
\begin{equation}
\mathcal{A}=\mathcal{Q},
\qquad\mathcal{H}=\mathcal{F}\otimes \mathbb{C}^4,
\end{equation}
and an element $e\in \mathcal{A}$ acts on
$\Psi=\left(|\psi_1\rangle,|\psi_2\rangle,|\psi_3\rangle,|\psi_4\rangle\right)'\in\mathcal{H}$
through the diagonal representation $\pi$ as
\begin{equation}
\pi(e)\Psi=\pi(e)\left(\begin{array}{c}
     |\psi_1\rangle \\
     |\psi_2\rangle \\
     |\psi_3\rangle \\
     |\psi_4\rangle \\
\end{array}\right)
=\left(\begin{array}{cccc}
e & 0 & 0 & 0 \\
0 & e & 0 & 0 \\
0 & 0 & e & 0 \\
0 & 0 & 0 & e \\
\end{array}\right)\left(\begin{array}{c}
     |\psi_1\rangle \\
     |\psi_2\rangle \\
     |\psi_3\rangle \\
     |\psi_4\rangle \\
\end{array}\right)
=\left(\begin{array}{c}
     e|\psi_1\rangle \\
     e|\psi_2\rangle \\
     e|\psi_3\rangle \\
     e|\psi_4\rangle \\
\end{array}\right).
\end{equation}
The Dirac operator for the 4D generalized noncommutative phase space (\ref{ncps}) can be defined as (see Appendix \ref{apa} for more details)
\begin{equation}\label{do}
\mathcal{D}=\beta\left(
  \begin{array}{cccc}
    0 & 0 & -\hat{A}_2^\dag & -\hat{A}_1^\dag \\
    0 & 0 &  \hat{A}_1 & -\hat{A}_2 \\
    -\hat{A}_2 & \hat{A}_1^\dag & 0 & 0 \\
    -\hat{A}_1 & -\hat{A}_2^\dag & 0 & 0 \\
  \end{array}\right),
\end{equation}
where
\begin{equation}\label{ba}
\beta=\frac{\sqrt{\hbar+\sqrt{\hbar^2-\theta^2}}}{\sqrt{\hbar^2-\theta^2}},
\qquad
\hat{A}_i=\hat{a}_i-\mathrm{i}t\varepsilon_{ij}\hat{a}_j,
\end{equation}
and
\begin{equation}
t=\frac{\theta}{\hbar+\sqrt{\hbar^2-\theta^2}},\qquad
\varepsilon=\left(
  \begin{array}{cc}
    0 & 1 \\
    -1 & 0 \\
  \end{array}
\right).
\end{equation}
Obviously, there is $0<t \ll 1$.

When the noncommutative parameters $\mu,\nu\to 0$, there is $\theta\to 0$, and $t\to 0$. The operators $\hat{A}_i$ will return to the operators $\hat{a}_i$, and the Dirac operator $\mathcal{D}$ (\ref{do}) will return to the Dirac operator of a normal 4D quantum phase space.

\section{Connes spectral distance and optimal elements}\label{sec3}
Let us consider the case where the quantum states $\omega$ are normal and bounded, so they are representable by density matrices $\rho$ \cite{Scholtz}.
The action of the state $\omega$ on an element $e\in \mathcal{A}$ can be written as
\begin{equation}
\omega(e)=\mathrm{tr}_{\mathcal{F}}(\rho e).
\end{equation}
Suppose the quantum states $\omega$ and $\omega'$ correspond to the density matrices $\rho$ and $\rho'$, respectively.
The Connes spectral distance between two states $\omega$ and $\omega'$ is \cite{Connes1}
\begin{equation}\label{d0}
d(\omega,\omega')\equiv \sup_{e\in B}|\omega(e)-\omega'(e)|
=\sup_{e\in B}|\mathrm{tr}_{\mathcal{F}}(\rho e)-\mathrm{tr}_{\mathcal{F}}(\rho' e)|,
\end{equation}
where
\begin{equation}\label{ball}
B=\left\{e\in \mathcal{A}:\big\|[\mathcal{D},\pi(e)]\big\|_{op}\leqslant 1\right\},
\end{equation}
and the operator norm is defined as
\begin{equation}
\|a\|_{op}\equiv\sup_{\psi\in \mathcal{H}}\frac{\|a\psi\|}{\|\psi\|},\qquad
\|a\|^2\equiv (a,a)=\mathrm{tr}_{\mathcal{F}}(a^{\dag}a).
\end{equation}
The elements $e$ which saturating the upper bound in (\ref{d0}) are the so-called optimal elements.

The inequality in (\ref{ball}) is the so-called ball condition.
Using the Dirac operator $\mathcal{D}$ (\ref{do}), the commutator $[\mathcal{D},\pi(e)]$ for a Hermitian element $e\in \mathcal{A}$ is
\begin{equation}
[\mathcal{D},\pi(e)]=\beta\left(
\begin{array}{cc}
    0 & D_1 \\
    -D_1^\dag & 0 \\
  \end{array}\right),
\end{equation}
where
\begin{equation}\label{d1d}
D_1=\left(
  \begin{array}{cc}
  [\hat{A}_2,e]^\dag & [\hat{A}_1,e]^\dag \\{}
  [\hat{A}_1,e] & -[\hat{A}_2,e] \\
  \end{array}\right).
\end{equation}
After some straightforward calculations, one can obtain
\begin{equation}\label{dd1}
\left\|[\mathcal{D},\pi(e)]^\dag[\mathcal{D},\pi(e)]\right\|_{op}
=\beta^2\max\left\{\|D_1^\dag D_1\|_{op},\,\|D_1 D_1^\dag\|_{op}\right\}.
\end{equation}
From the above expression (\ref{d1d}), we have
\begin{equation}\label{dd2}
D_1^\dag D_1
=\left(
  \begin{array}{cc}
  [\hat{A}_2,e][\hat{A}_2,e]^\dag+[\hat{A}_1,e]^\dag[\hat{A}_1,e] & ~[\hat{A}_2,e][\hat{A}_1,e]^\dag-[\hat{A}_1,e]^\dag[\hat{A}_2,e] \\{}
  [\hat{A}_1,e][\hat{A}_2,e]^\dag-[\hat{A}_2,e]^\dag[\hat{A}_1,e] & ~[\hat{A}_1,e][\hat{A}_1,e]^\dag+[\hat{A}_2,e]^\dag[\hat{A}_2,e] \\
  \end{array}\right).
\end{equation}
For any operator $M$ with the matrix elements $M_{ij}$ in some orthonormal bases, there is the following Bessel's inequality \cite{Revisiting},
\begin{equation}\label{bi}
\|M\|_{op}^2\geqslant \sum_i|M_{ij}|^2\geqslant |M_{ij}|^2.
\end{equation}
So there should be
\begin{equation}\label{d1d1}
\|D_1^\dag D_1\|_{op}
\geqslant\!\sup_{\phi\in\mathcal{F},\langle\phi|\phi\rangle=1}
\!\!\!\!\!\!\langle\phi|[\hat{A}_1,e]^\dag[\hat{A}_1,e]+[\hat{A}_2,e][\hat{A}_2,e]^\dag|\phi\rangle.
\end{equation}
It is known that for any bounded operator $M$, there is
$\|M\|_{op}^2=\|M^\dag\|_{op}^2=\|M^\dag M\|_{op}$.
Using the ball condition in (\ref{ball}), there is the following inequality,
\begin{equation}\label{bc}
1 \geqslant \big\|[\mathcal{D},\pi(e)]\big\|_{op}^2
=\left\|[\mathcal{D},\pi(e)]^\dag[\mathcal{D},\pi(e)]\right\|_{op}
=\beta^2\|D_1^\dag D_1\|_{op}.
\end{equation}
So from (\ref{d1d1}) and (\ref{bc}), for any Hermitian element $e\in B$, one can obtain
\begin{equation}\label{bc1}
\sup_{\phi\in\mathcal{F},\langle\phi|\phi\rangle=1}
\!\!\!\!\!\!\langle\phi|[\hat{A}_1,e]^\dag[\hat{A}_1,e]+[\hat{A}_2,e][\hat{A}_2,e]^\dag|\phi\rangle
\leqslant\frac{1}{\beta^2}.
\end{equation}
Similarly, we also have the following inequalities,
\begin{eqnarray}\label{bc2}
&&\sup_{\phi\in\mathcal{F},\langle\phi|\phi\rangle=1}
\!\!\!\!\!\!\langle\phi|[\hat{A}_1,e][\hat{A}_1,e]^\dag+[\hat{A}_2,e]^\dag[\hat{A}_2,e]|\phi\rangle
\leqslant\frac{1}{\beta^2},\nonumber\\
&&\sup_{\phi\in\mathcal{F},\langle\phi|\phi\rangle=1}
\!\!\!\!\!\!\langle\phi|[\hat{A}_1,e]^\dag[\hat{A}_1,e]+[\hat{A}_2,e]^\dag[\hat{A}_2,e]|\phi\rangle
\leqslant\frac{1}{\beta^2},
\nonumber\\
&&\sup_{\phi\in\mathcal{F},\langle\phi|\phi\rangle=1}
\!\!\!\!\!\!\langle\phi|[\hat{A}_1,e][\hat{A}_1,e]^\dag+[\hat{A}_2,e][\hat{A}_2,e]^\dag|\phi\rangle
\leqslant\frac{1}{\beta^2}.
\end{eqnarray}

Now consider an element $e\in B$ which can be expressed as
\begin{equation}\label{eee}
e=\sum_{i,j,k,l=0}^\infty C^{i,j}_{k,l}|i,j\rangle\langle k,l|.
\end{equation}
Since Hermitian elements can give the supremum in the Connes spectral
distance functions \cite{Iochum}, we only need to consider the element $e$ being Hermitian. So there are $C^{i,j}_{k,l}=(C^{k,l}_{i,j})^*$, and $C^{i,j}_{i,j}$ are real numbers.
Using the relations (\ref{a1a2}), after some straightforward calculations, we have
\begin{eqnarray}
[\hat{A}_1,e]&=&[\hat{a}_1-\mathrm{i}t\hat{a}_2,e]\nonumber\\
&=&\sum_{i,j,k,l=0}^\infty\bigg(\left(C^{i+1,j}_{k,l}\sqrt{i+1}
-C^{i,j}_{k-1,l}\sqrt{k}\right)
-\mathrm{i}t\left(C^{i,j+1}_{k,l}\sqrt{j+1}
-C^{i,j}_{k,l-1}\sqrt{l}\right)\bigg)|i,j\rangle\langle k,l|,
\end{eqnarray}
and then
\begin{eqnarray}\label{a1e}
[\hat{A}_1,e][\hat{A}_1,e]^\dag&=&
\sum_{i,j=0}^\infty\Bigg(\sum_{k,l=0}^\infty\bigg|
\left(C^{i+1,j}_{k,l}\sqrt{i+1}
-C^{i,j}_{k-1,l}\sqrt{k}\right)\nonumber\\
&&\quad-\mathrm{i}t\left(C^{i,j+1}_{k,l}\sqrt{j+1}
-C^{i,j}_{k,l-1}\sqrt{l}\right)
\bigg|^2\Bigg)|i,j\rangle\langle i,j|+\text{n.d.}\,.
\end{eqnarray}
Here ``n.d.'' denotes the sum of terms $|k,l\rangle\langle i,j|$ with $k\neq i$ and/or $l\neq j$, and these terms will not affect the calculations in the following content. So we can just simply ignore these terms.

For simplicity, we will consider the optimal elements being diagonal elements first. In the following, we will show how to find some optimal elements in diagonal elements. Therefore we can set $C^{i,j}_{k,l}=0$ if $k\neq i$ and/or $l\neq j$, and denote $c_{i,j}\equiv C^{i,j}_{i,j}$.
For convenience, we also denote
\begin{equation}
E_{i,j}\equiv c_{i,j}-c_{i-1,j},\qquad
F_{i,j}\equiv c_{i,j}-c_{i,j-1},
\end{equation}
and
\begin{equation}
G_{i,j}\equiv iE_{i,j}^2=i(c_{i,j}-c_{i-1,j})^2,\qquad
H_{i,j}\equiv jF_{i,j}^2=j(c_{i,j}-c_{i,j-1})^2.
\end{equation}
So the above equation (\ref{a1e}) can be rewritten as
\begin{eqnarray}\label{ae11}
[\hat{A}_1,e][\hat{A}_1,e]^\dag&=&
\sum_{i,j=0}^\infty\big((i+1)(c_{i+1,j}-c_{i,j})^2
+t^2(j+1)(c_{i,j+1}-c_{i,j})^2\big)|i,j\rangle\langle i,j|+\text{n.d.}\nonumber\\
&=&\sum_{i,j=0}^\infty\left((i+1)E_{i+1,j}^2
+t^2(j+1)F_{i,j+1}^2\right)|i,j\rangle\langle i,j|+\text{n.d.}\nonumber\\
&=&\sum_{i,j=0}^\infty\left(G_{i+1,j}
+t^2H_{i,j+1}\right)|i,j\rangle\langle i,j|+\text{n.d.}\,.
\end{eqnarray}
Similarly, there is
\begin{equation}\label{ae12}
[\hat{A}_1,e]^\dag[\hat{A}_1,e]
=\sum_{i,j=0}^\infty\big(G_{i,j}+t^2 H_{i,j}
\big)|i,j\rangle\langle i,j|+\text{n.d.}\,.
\end{equation}

For the operator $\hat{A}_2$, we also have
\begin{eqnarray}
[\hat{A}_2,e]
&=&\sum_{i,j,k,l=0}^\infty\bigg(\left(C^{i,j+1}_{k,l}\sqrt{j+1}
-C^{i,j}_{k,l-1}\sqrt{l}\right)
+\mathrm{i}t\left(C^{i+1,j}_{k,l}\sqrt{i+1}
-C^{i,j}_{k-1,l}\sqrt{k}\right)\bigg)|i,j\rangle\langle k,l|,
\end{eqnarray}
and
\begin{eqnarray}\label{ae21}
[\hat{A}_2,e][\hat{A}_2,e]^\dag
&=&\sum_{i,j=0}^\infty\big(H_{i,j+1}
+t^2G_{i+1,j}\big)|i,j\rangle\langle i,j|+\text{n.d.}\,,\nonumber\\{}
[\hat{A}_2,e]^\dag[\hat{A}_2,e]
&=&\sum_{i,j=0}^\infty\big(H_{i,j}
+t^2 G_{i,j}\big)|i,j\rangle\langle i,j|+\text{n.d.}\,.
\end{eqnarray}
From the inequalities (\ref{bc1}) and (\ref{bc2}), using the results (\ref{ae11}), (\ref{ae12}) and (\ref{ae21}), one can obtain the following relations,
\begin{eqnarray}\label{cr}
&&(G_{i+1,j}+H_{i,j})+t^2(G_{i,j}+H_{i,j+1})\leqslant\frac{1}{\beta^2},\quad
(1+t^2)(G_{i+1,j}+H_{i,j+1})\leqslant\frac{1}{\beta^2},\nonumber\\
&&(G_{i,j}+H_{i,j+1})+t^2(G_{i+1,j}+H_{i,j})\leqslant\frac{1}{\beta^2},\quad (1+t^2)(G_{i,j}+H_{i,j})\leqslant\frac{1}{\beta^2}.\qquad\quad
\end{eqnarray}

So in the following calculations of Connes spectral distances, we will consider the optimal elements being diagonal,
\begin{equation}\label{eo0}
e_o=\sum_{i,j=0}^\infty c_{i,j}|i,j\rangle\langle i,j|,
\end{equation}
where the coefficients $c_{i,j}$ satisfy the above constraint relations (\ref{cr}). The Connes spectral distance between the quantum states
$\omega$ and $\omega'$ is obtained by
\begin{equation}
d(\omega,\omega')
=|\mathrm{tr}_{\mathcal{F}}(\rho\, e_o)-\mathrm{tr}_{\mathcal{F}}(\rho'\, e_o)|.
\end{equation}

\section{Connes spectral distance between Fock states in NCPS}\label{sec4}
Using the spectral triple $(\mathcal{A},\mathcal{H},\mathcal{D})$ constructed above, one can calculate the Connes spectral distances between Fock states $|n_1,n_2\rangle$.
First, let us consider the adjacent Fock states $|m+1,n\rangle$ and $|m,n\rangle$, and the corresponding density matrices are $\rho_{m+1,n}=|m+1,n\rangle\langle m+1,n|$ and $\rho_{m,n}=|m,n\rangle\langle m,n|$, respectively. The spectral distance is
\begin{eqnarray}\label{d3}
d(\omega_{m+1,n},\omega_{m,n})&=&\sup_{e\in B}|\mathrm{tr}_\mathcal{F}(\rho_{m+1,n} e)-\mathrm{tr}_\mathcal{F}(\rho_{m,n} e)|\nonumber\\
&=&\sup_{e\in B}|\langle m+1,n|e|m+1,n\rangle-\langle m,n|e|m,n\rangle|\nonumber\\
&=&\sup_{e\in B}\frac{1}{\sqrt{m+1}}|\langle m,n|[\hat{a}_1,e]|m+1,n\rangle|\nonumber\\
&\leqslant&\frac{1}{\sqrt{m+1}}\big\|[\hat{a}_1,e]\big\|_{op},
\end{eqnarray}
and we have used the Bessel's inequality (\ref{bi}) in the above inequality.

From (\ref{ba}), after some straightforward calculations, one can obtain
\begin{equation}
[\hat{A}_1,e]^\dag[\hat{A}_1,e]+[\hat{A}_2,e]^\dag[\hat{A}_2,e]
=(1+t^2)\big([\hat{a}_1,e]^\dag[\hat{a}_1,e]+[\hat{a}_2,e]^\dag[\hat{a}_2,e]\big)
+2\mathrm{i}t[\hat{a}_2,e]^\dag[\hat{a}_1,e]-2\mathrm{i}t[\hat{a}_1,e]^\dag[\hat{a}_2,e].
\end{equation}
Since the operators $\hat{a}_1$ and $\hat{a}_2$ commute
with each other, which correspond to two independent $1D$ harmonic oscillators.
So it is reasonable to conjecture that the Connes distance between the states $|m,n\rangle$ and $|m+1,n\rangle$ does not depend on the second parameter $n$. This should be similar to the Connes distance between the $1D$ harmonic oscillator states $|m\rangle$ and $|m+1\rangle$. Therefore the corresponding optimal elements $e$ in this case can be chosen to commute with $\hat{a}_2$, $[\hat{a}_2,e]=0$.
So from (\ref{bc2}), for the optimal elements corresponding to the Connes distance between the states $|m,n\rangle$ and $|m+1,n\rangle$, there should be
\begin{equation}
\sup_{\phi\in\mathcal{F},\langle\phi|\phi\rangle=1}
\!\!\!\!\!\!\langle\phi|[\hat{a}_1,e]^\dag[\hat{a}_1,e]|\phi\rangle
\leqslant\frac{1}{\beta^2(1+t^2)},
\end{equation}
or
\begin{equation}\label{a1a1d}
\big\|[\hat{a}_1,e]\big\|_{op}^2
\leqslant\frac{1}{\beta^2(1+t^2)}.
\end{equation}
Combining the results (\ref{d3}) and (\ref{a1a1d}), we have
\begin{equation}
d(\omega_{m+1,n},\omega_{m,n})
\leqslant\frac{1}{\sqrt{m+1}}\frac{1}{\beta\sqrt{1+t^2}}.
\end{equation}

In the following, we will use the constraint relations (\ref{cr}) to find some optimal elements in the diagonal elements, and they can saturate the above inequality. Using the expression (\ref{eo0}), we also have
\begin{eqnarray}
d(\omega_{m+1,n},\omega_{m,n})&=&
\sup_{e\in B}|\langle m+1,n|e|m+1,n\rangle-\langle m,n|e|m,n\rangle|\nonumber\\
&=&\sup_{e\in B}|c_{m+1,n}-c_{m,n}|
=\sup_{e\in B}|E_{m+1,n}|.
\end{eqnarray}

From (\ref{cr}), we have
\begin{eqnarray}\label{ghgh1}
&&(1+t^2)(G_{m+1,n}+H_{m,n+1})\leqslant\frac{1}{\beta^2},\qquad
(1+t^2)(G_{m+1,n}+H_{m+1,n})\leqslant\frac{1}{\beta^2},\nonumber\\
&& (G_{m+1,n}+H_{m+1,n+1})+t^2(G_{m+2,n}+H_{m+1,n})\leqslant\frac{1}{\beta^2},\nonumber\\
&& (G_{m+1,n}+H_{m,n})+t^2(G_{m,n}+H_{m,n+1})\leqslant\frac{1}{\beta^2}.
\end{eqnarray}
So in order to attain the supremum of $|c_{m+1,n}-c_{m,n}|=|E_{m+1,n}|$, there should be
\begin{equation}
H_{m,n+1}=H_{m+1,n}=0,
\end{equation}
and
\begin{equation}\label{me}
(1+t^2)G_{m+1,n}=(1+t^2)(m+1)E_{m+1,n}^2=\frac{1}{\beta^2},
\end{equation}
or
\begin{equation}
|c_{m+1,n}-c_{m,n}|=|E_{m+1,n}|=\frac{1}{\beta\sqrt{(1+t^2)(m+1)}}.
\end{equation}
We can also set
\begin{equation}
H_{m,n}=H_{m+1,n+1}=0,
\end{equation}
so there is
\begin{equation}
F_{m,n+1}=F_{m+1,n}=F_{m,n}=F_{m+1,n+1}=0.
\end{equation}
Without loss of generality, we can assume $E_{m+1,n}\geqslant 0$. There is
\begin{eqnarray}
E_{m+1,n}&=&F_{m+1,n+1}+E_{m+1,n}=c_{m+1,n+1}-c_{m,n}\nonumber\\
&=&E_{m+1,n+1}+F_{m,n+1}=E_{m+1,n+1}.
\end{eqnarray}
Similarly, one can obtain
\begin{equation}
E_{m+1,i}=\frac{1}{\beta\sqrt{(1+t^2)(m+1)}},\qquad
F_{m,i+1}=F_{m+1,i+1}=0,
\end{equation}
where $i=0,1,2,...$\,.

For example, one can choose
\begin{equation}
c_{m,i}=0\,,\qquad
c_{m+1,i}=\frac{1}{\beta\sqrt{(1+t^2)(m+1)}}\,,
\end{equation}
and the optimal element is
\begin{eqnarray}
e_o&=&\sum_{i=0}^\infty
c_{m+1,i}|m+1,i\rangle\langle m+1,i|\nonumber\\
&=&\frac{1}{\beta\sqrt{(1+t^2)(m+1)}}
\sum_{i=0}^\infty|m+1,i\rangle\langle m+1,i|\nonumber\\
&=&\frac{1}{\beta\sqrt{(1+t^2)(m+1)}}
|m+1\rangle_{\!1\,1\!}\langle m+1|\otimes\mathds{I}_2.
\end{eqnarray}
So the Connes distance between the states $|m+1,n\rangle$ and $|m,n\rangle$ is
\begin{eqnarray}
d(\omega_{m+1,n},\omega_{m,n})
&=&|\mathrm{tr}_{\mathcal{F}}(\rho_{m+1,n} e_o)-\mathrm{tr}_{\mathcal{F}}(\rho_{m,n} e_o)|\nonumber\\
&=&|\langle m+1,n|e_o|m+1,n\rangle-\langle m,n|e_o|m,n\rangle|\nonumber\\
&=&\frac{1}{\beta\sqrt{(1+t^2)(m+1)}}\nonumber\\
&=&\frac{\sqrt{\hbar^2-\theta^2}}{\sqrt{2\hbar}}\frac{1}{\sqrt{m+1}}.
\end{eqnarray}

Similarly, the Connes spectral distance between Fock states $|m,n+1\rangle$ and $|m,n\rangle$ is
\begin{equation}
d(\omega_{m,n+1},\omega_{m,n})
=\frac{1}{\beta\sqrt{(1+t^2)(n+1)}}
=\frac{\sqrt{\hbar^2-\theta^2}}{\sqrt{2\hbar}}\frac{1}{\sqrt{n+1}}.
\end{equation}

Next, let us consider the Fock states $|m,n\rangle$ and $|m+k,n\rangle$, $k>1$. The corresponding Connes spectral distance is
\begin{eqnarray}
d(\omega_{m+k,n},\omega_{m,n})&=&\sup_{e\in B}|\mathrm{tr}_\mathcal{F}(\rho_{m+k,n} e)-\mathrm{tr}_\mathcal{F}(\rho_{m,n} e)|\nonumber\\
&=&\sup_{e\in B}|\mathrm{tr}_\mathcal{F}(\rho_{m+k,n} e)-\mathrm{tr}_\mathcal{F}(\rho_{m+k-1,n} e)
+\mathrm{tr}_\mathcal{F}(\rho_{m+k-1,n} e)-\mathrm{tr}_\mathcal{F}(\rho_{m+k-2,n} e)\nonumber\\
&&+...+\mathrm{tr}_\mathcal{F}(\rho_{m+1,n} e)-\mathrm{tr}_\mathcal{F}(\rho_{m,n} e)|\nonumber\\
&\leqslant&
\sup_{e\in B}\sum_{i=1}^{k}|\mathrm{tr}_\mathcal{F}(\rho_{m+i,n} e)-\mathrm{tr}_\mathcal{F}(\rho_{m+i-1,n} e)|\nonumber\\
&=&\sup_{e\in B}\sum_{i=1}^{k}\frac{1}{\sqrt{m+i}}|\langle m+i-1,n|[\hat{a}_1,e]|m+i,n\rangle|\nonumber\\
&\leqslant&\sum_{i=1}^{k}\frac{1}{\sqrt{m+i}}\big\|[\hat{a}_1,e]\big\|_{op}\nonumber\\
&\leqslant&\frac{1}{\beta\sqrt{1+t^2}}\sum_{i=1}^{k}\frac{1}{\sqrt{m+i}}.
\end{eqnarray}
Using the expression (\ref{eo0}), the distance $d(\omega_{m+k,n},\omega_{m,n})$ can also be expressed as
\begin{eqnarray}\label{d1}
d(\omega_{m+k,n},\omega_{m,n})
&=&\sup_{e\in B}|c_{m+k,n}-c_{m,n}|\nonumber\\
&=&\sup_{e\in B}|c_{m+k,n}-c_{m+k-1,n}+c_{m+k-1,n}-c_{m+k-2,n}\nonumber\\
&&\qquad+...+c_{m+1,n}-c_{m,n}|\nonumber\\
&=&\sup_{e\in B}|E_{m+k,n}+E_{m+k-1,n}+...+E_{m+1,n}|.
\end{eqnarray}
From (\ref{cr}), we have
\begin{eqnarray}
&&(1+t^2)(G_{m+2,n}+H_{m+1,n+1})\leqslant\frac{1}{\beta^2},\qquad
(1+t^2)(G_{m+1,n}+H_{m+1,n})\leqslant\frac{1}{\beta^2},\nonumber\\
&& (G_{m+2,n}+H_{m+1,n})+t^2(G_{m+1,n}+H_{m+1,n+1})\leqslant\frac{1}{\beta^2},\nonumber\\
&& (G_{m+1,n}+H_{m+1,n+1})+t^2(G_{m+2,n}+H_{m+1,n})\leqslant\frac{1}{\beta^2}.
\end{eqnarray}
So in order to attain the supremum of $|c_{m+k,n}-c_{m,n}|=|E_{m+k,n}+E_{m+k-1,n}+...+E_{m+1,n}|$,
one must let $|c_{m+2,n}-c_{m,n}|=|E_{m+2,n}+E_{m+1,n}|$ as large as possible.
There should be $H_{m+1,n}=H_{m+1,n+1}=0$, and
\begin{eqnarray}
&& G_{m+2,n}+t^2G_{m+1,n}=\frac{1}{\beta^2},\qquad G_{m+1,n}+t^2G_{m+2,n}=\frac{1}{\beta^2},\nonumber\\
&&(1+t^2)G_{m+2,n}=\frac{1}{\beta^2},\qquad
(1+t^2)G_{m+1,n}=\frac{1}{\beta^2}.
\end{eqnarray}
Therefore we have
\begin{equation}
G_{m+1,n}=G_{m+2,n}=\frac{1}{\beta^2(1+t^2)}.
\end{equation}

Similarly, in order to attain the supremum of $|c_{m+k,n}-c_{m,n}|$, there must be
\begin{equation}\label{cce}
H_{m+i,n}=H_{m+i,n+1}=0,\qquad
G_{m+i,n}=\frac{1}{\beta^2(1+t^2)},\qquad i=1,2,...,k\,,
\end{equation}
so
\begin{equation}
c_{m+i,n}-c_{m+i-1,n}=E_{m+i,n}=\frac{1}{\beta\sqrt{1+t^2}}\frac{1}{\sqrt{m+i}}.
\end{equation}
For example, one can choose
\begin{equation}\label{ccc}
c_{m+p,i}
=\frac{1}{\beta\sqrt{1+t^2}}
\sum_{j=1}^{m+p}\frac{1}{\sqrt{j}}=\zeta_{0;m+p}\frac{1}{\beta\sqrt{1+t^2}},
\end{equation}
where $i=0,1,2...$\,, $p=0,1,...,k$, and the function $\zeta_{p;q}$ is
\begin{equation}
\zeta_{p;q}\equiv\sum_{i=p+1}^{q}\frac{1}{\sqrt{i}}
=\zeta\left(\frac{1}{2},p+1\right)
-\zeta\left(\frac{1}{2},q+1\right),
\end{equation}
where $\zeta(s,q)$ is the Hurwitz zeta function
\begin{equation}
\zeta(s,q)\equiv\sum_{i=0}^{\infty}\frac{1}{(q+i)^s}.
\end{equation}
Obviously, there is $\zeta_{i;j}=-\zeta_{j;i}$, and therefore $\zeta_{i;i}=0$.
We also have $\zeta_{i;j}-\zeta_{i;k}=\zeta_{k;j}$.
It is easy to verify that these coefficients $c_{i,j}$ (\ref{ccc}) satisfy the relations (\ref{cce}).

Suppose $e_o$ to be the optimal element with the coefficients $c_{i,j}$ (\ref{ccc}), it can attain the supremum in (\ref{d1}). There is
\begin{eqnarray}\label{eo11}
e_o&=&\sum_{p=0}^{k}\sum_{i=0}^\infty
c_{m+p,i}|m+p,i\rangle\langle m+p,i|
\nonumber\\
&=&\frac{1}{\beta\sqrt{1+t^2}}\sum_{p=0}^{k}\zeta_{0;m+p}
|m+p\rangle_{\!1\,1\!}\langle m+p|\otimes\mathds{I}_2,
\end{eqnarray}
and the Connes spectral distance between the states $|m+k,n\rangle$ and $|m,n\rangle$ is
\begin{eqnarray}\label{d11}
d(\omega_{m+k,n},\omega_{m,n})&=&|\mathrm{tr}_{\mathcal{F}}(\rho_{m+k,n} e_o)-\mathrm{tr}_{\mathcal{F}}(\rho_{m,n} e_o)|\nonumber\\
&=&|\langle m+k,n|e_o|m+k,n\rangle-\langle m,n|e_o|m,n\rangle|\nonumber\\
&=&\frac{1}{\beta\sqrt{1+t^2}}
\sum_{i=1}^{k}\frac{1}{\sqrt{m+i}}
=\frac{\sqrt{\hbar^2-\theta^2}}{\sqrt{2\hbar}}
\sum_{i=1}^{k}\frac{1}{\sqrt{m+i}}\nonumber\\
&=&\zeta_{m;m+k}\frac{\sqrt{\hbar^2-\theta^2}}{\sqrt{2\hbar}}.
\end{eqnarray}
Obviously, there is $d(\omega_{m+k,n},\omega_{m,n})=d(\omega_{m,n},\omega_{m+k,n})$.
These distances do not depend on $n$, for any integers $n\neq n'$, there should be $d(\omega_{m+k,n},\omega_{m,n})=d(\omega_{m+k,n'},\omega_{m,n'})$.

It is easy to see that, for $m,n,k,l\geqslant 0$, these distances satisfy
\begin{equation}
d(\omega_{m+k+l,n},\omega_{m,n})
=d(\omega_{m+k,n},\omega_{m,n})+d(\omega_{m+k+l,n},\omega_{m+k,n}).
\end{equation}
This means that these distances are additive.

Similarly, the Connes spectral distance between Fock states $|m,n+k\rangle$ and $|m,n\rangle$ is
\begin{equation}
d(\omega_{m,n+k},\omega_{m,n})
=\frac{\sqrt{\hbar^2-\theta^2}}{\sqrt{2\hbar}}
\sum_{i=1}^{k}\frac{1}{\sqrt{n+i}}
=\zeta_{n;n+k}\frac{\sqrt{\hbar^2-\theta^2}}{\sqrt{2\hbar}}.
\end{equation}
We also have $d(\omega_{m,n+k},\omega_{m,n})=d(\omega_{m',n+k},\omega_{m',n})$ for $m\neq m'$.

Compare with the result in Refs.~\cite{Cagnache,Scholtz,Revisiting}, one can find that the Connes spectral distances between Fock states in the generalized noncommutative phase space have one more factor $\sqrt{1-\theta^2\!/\hbar^2}=\sqrt{1-\mu\nu\!/\hbar^2}$ than those in normal quantum phase space. So in general, the Connes spectral distances in generalized noncommutative phase spaces are shorter than those in normal quantum phase spaces. This is intuitive and reasonable, because the noncommutativity of the space will lead to some type of nonlocality. The partial nonlocality can equivalently make the distances shorter.

Obviously, when $\theta\to \hbar$, these Connes spectral distances will become zero. This will cause some singularities.

Finally, let us study the Connes distance between Fock states $|m,n\rangle$ and $|m+k,n+l\rangle$. For simplicity, we assume $k,l>1$. Using the expression (\ref{eee}), we have
\begin{eqnarray}
d(\omega_{m+k,n+l},\omega_{m,n})
&=&\sup_{e\in B}|\mathrm{tr}_{\mathcal{F}}(\rho_{m+k,n+l}\,e)-\mathrm{tr}_{\mathcal{F}}(\rho_{m,n}\,e)|
=\sup_{e\in B}|c_{m+k,n+l}-c_{m,n}|\nonumber\\
&=&\sup_{e\in B}|\mathrm{tr}_{\mathcal{F}}(\rho_{m+k,n+l}\,e)-\mathrm{tr}_{\mathcal{F}}(\rho_{m+k,n}\,e)
+\mathrm{tr}_{\mathcal{F}}(\rho_{m+k,n}\,e)-\mathrm{tr}_{\mathcal{F}}(\rho_{m,n}\,e)|\nonumber\\
&=&\sup_{e\in B}|c_{m+k,n+l}-c_{m+k,n}+c_{m+k,n}-c_{m,n}|.
\end{eqnarray}
Without loss of generality, we can assume $c_{m+k,n+l}\geqslant c_{m+k,n} \geqslant c_{m,n}$. So in order to attain the supremum of $|c_{m+k,n+l}-c_{m,n}|$,
one must let $c_{m+k,n+l}-c_{m+k,n}$ and also $c_{m+k,n}-c_{m,n}$ as large as possible. This is equivalent to choose one element $e$ to make $|\mathrm{tr}_{\mathcal{F}}(\rho_{m+k,n+l}\,e)-\mathrm{tr}_{\mathcal{F}}(\rho_{m+k,n}\,e)|$ and $|\mathrm{tr}_{\mathcal{F}}(\rho_{m+k,n}\,e)-\mathrm{tr}_{\mathcal{F}}(\rho_{m,n}\,e)|$
as large as possible.
Similarly, one can see that, for any $0\leqslant i \leqslant l$, $0\leqslant j \leqslant k$ and $0\leqslant i_1 \leqslant i_2 \leqslant l$, $0\leqslant j_1 \leqslant j_2 \leqslant k$, the element $e$ should make
\begin{eqnarray}
c_{m+j,n+i_2}-c_{m+j,n+i_1}&=&\mathrm{tr}_{\mathcal{F}}(\rho_{m+j,n+i_2}\,e)-\mathrm{tr}_{\mathcal{F}}(\rho_{m+j,n+i_1}\,e),\nonumber\\ c_{m+j_2,n+i}-c_{m+j_1,n+i}&=&\mathrm{tr}_{\mathcal{F}}(\rho_{m+j_2,n+i}\,e)-\mathrm{tr}_{\mathcal{F}}(\rho_{m+j_1,n+i}\,e)
\end{eqnarray}
as large as possible.
These are similar to the calculations of the Connes distance between $|m+j,n+i_2\rangle$, $|m+j,n+i_1\rangle$ and that between $|m+j_2,n+i\rangle$, $|m+j_1,n+i\rangle$, respectively.

So comparing with the expression (\ref{eo11}), one can assume that the corresponding optimal element $e_o$ contains the following terms,
\begin{equation}
|m+j\rangle_{\!1\,1\!}\langle m+j|\otimes
\sum_{q=n+i_1}^{n+i_2}\!\!\!\!\lambda_2\zeta_{0;q}
|q\rangle_{\!2\,2\!}\langle q|,
\qquad
\sum_{p=m+j_1}^{m+j_2}\!\!\!\!\lambda_1\zeta_{0;p}
|p\rangle_{\!1\,1\!}\langle p|\otimes|n+i\rangle_{\!2\,2\!}\langle n+i|,
\end{equation}
where $\lambda_1$, $\lambda_2$ are some constants. These terms are all diagonal.
Furthermore, similar to the calculations of the Connes distance between $|m+k,n\rangle$ and $|m,n\rangle$, this also implies that it is possible to construct some optimal elements which satisfy the above conditions in diagonal elements.
So one can still using the constraint relations (\ref{cr}) for diagonal elements
to construct the optimal elements. For example, one can choose the following optimal element (see Appendix \ref{apb} for more details),
\begin{equation}
e_o=\frac{1}{\beta\sqrt{1+t^2}}\sum_{p=m}^{m+k}\sum_{q=n}^{n+l}
\frac{\zeta_{0;p}\zeta_{m;m+k}+\zeta_{0;q}\zeta_{n;n+l}}{\sqrt{\zeta_{m;m+k}^2
+\zeta_{n;n+l}^2}}|p,q\rangle\langle p,q|,
\end{equation}
and the Connes spectral distance between the Fock states $|m+k,n+l\rangle$ and $|m,n\rangle$ is
\begin{equation}
d(\omega_{m+k,n+l},\omega_{m,n})
=\frac{1}{\beta\sqrt{1+t^2}}
\sqrt{\zeta_{m;m+k}^2+\zeta_{n;n+l}^2}\,.
\end{equation}
These results are also true for $k\leqslant0$ and/or $l\leqslant0$.

It is easy to verify that the Connes spectral distances between Fock states also satisfy the Pythagoras theorem,
\begin{eqnarray}
d(\omega_{m+k,n+l},\omega_{m,n})
&=&\sqrt{d(\omega_{m+k,n+l},\omega_{m+k,n})^2+d(\omega_{m+k,n+l},\omega_{m,n+l})^2}
\nonumber\\
&=&\sqrt{d(\omega_{m+k,n+l},\omega_{m,n+l})^2+d(\omega_{m,n+l},\omega_{m,n})^2}\nonumber\\
&=&\sqrt{d(\omega_{m+k,n+l},\omega_{m+k,n})^2+d(\omega_{m+k,n},\omega_{m,n})^2}
\nonumber\\
&=&\sqrt{d(\omega_{m+k,n},\omega_{m,n})^2+d(\omega_{m,n+l},\omega_{m,n})^2}.
\end{eqnarray}
This is similar to the result obtained in Ref.~\cite{DAndrea}.

\section{Discussions and conclusions}\label{sec5}
In this paper, we study the Connes spectral distance of Fock states in a 4D generalized noncommutative phase space. By virtue of the Hilbert-Schmidt operatorial formulation, we obtain the Dirac operator and a spectral triple corresponding to the generalized noncommutative phase space. Based on the ball condition, we obtain some constraint relations about the optimal elements. Using these constraint relations, we construct the corresponding optimal elements and derive the Connes spectral distance between two Fock states.

We find that the spectral distances in noncommutative phase spaces are shorter than those in normal phase spaces. This is intuitive and reasonable, because the noncommutativity of the space will lead to some type of nonlocality.
This nonlocality can make the distances shorter equivalently.
These spectral distances in the 4D generalized noncommutative phase space are additive and satisfy the normal Pythagoras theorem. When the noncommutative parameter $\theta=\sqrt{\mu\nu}$ goes to zero, the results return to those in normal quantum phase space. But if $\theta\to \hbar$, these Connes spectral distances will go to zero. This may cause some singularities.

Here we only study the Connes spectral distances between Fock states. Obviously, these distances depend on the specific quantum states being considered.
One can also study the spectral distances between other kinds of pure states and mixed states.
In a noncommutative space, a pure state is the analog of a traditional point in a normal commutative space, and the spectral distance between pure states corresponds to the geodesic distance between points. Different kinds of quantum states can form different kinds of abstract spaces, and these spaces have different mathematical structures.
So studies of the Connes spectral distances between quantum states can help us to study the mathematical structures of the spaces and also physical properties of the quantum systems.

Our results are significant for the study of the Connes spectral distances of physical systems in noncommutative spaces.
Our methods can be used to study other physical systems in other kinds of noncommutative spaces.
Using our method, one can also study the Connes spectral distance between quantum states in higher-dimensional noncommutative spaces.
But usually the calculations and results will be much more complicated.

\section*{Acknowledgements}
The authors would like to thank the anonymous reviewer for the careful reading and constructive comments.
This work is partly supported by Key Research and Development Project of Guangdong Province (Grant No. 2020B0303300001), the Guangdong Basic and Applied Basic Research Foundation (Grant No. 2019A1515011703), the Fundamental Research Funds for the Central Universities and the Natural Science Foundation of Anhui Province (Grant No. 1908085MA16) and National Natural Science Foundation of China (Grant No. 11911530750).

\appendix

\section{Dirac operator for 4D generalized noncommutative phase space}\label{apa}
Similar to Ref.~\cite{Revisiting}, in order to construct the Dirac operator for the 4D generalized noncommutative phase space (\ref{ncps}), one can consider the following extended noncommutative phase space in which the coordinate operators $\tilde{X}_i$, $\tilde{Y}_i$ and $\tilde{P}_i$, $\tilde{Q}_i$ satisfy the following commutation relations,
\begin{eqnarray}\label{enc}
&&[\tilde{X}_i,\,\tilde{P}_j]=[\tilde{Y}_i,\,\tilde{Q}_j]
=\mathrm{i}\delta_{ij}\hbar,\qquad
[\tilde{X}_i,\,\tilde{Y}_j]=\mathrm{i}\delta_{ij}\lambda,\qquad
[\tilde{Q}_i,\,\tilde{P}_j]=\mathrm{i}\delta_{ij}\lambda,
\nonumber\\
&&[\tilde{X}_1,\,\tilde{X}_2]=[\tilde{Y}_1,\,\tilde{Y}_2]=\mathrm{i}\theta,\qquad
[\tilde{P}_1,\,\tilde{P}_2]=[\tilde{Q}_1,\,\tilde{Q}_2]=\mathrm{i}\theta,
\end{eqnarray}
where $\lambda$ is some parameter.

It is easy to verify that, the following unitary representation acting on the quantum Hilbert space $\mathcal{Q}$ satisfies the above noncommutative relations (\ref{enc}),
\begin{eqnarray}
&&\tilde{X}_i|\phi)
=\alpha\big|(\tilde{x}_i-t\varepsilon_{ij}\tilde{p}_j)\phi\big)\,,\qquad
\tilde{P}_i|\phi)
=\alpha\big|(\tilde{p}_i+t\varepsilon_{ij}\tilde{x}_j)\phi\big)\,,\nonumber\\
&&\tilde{Y}_i|\phi)
=\frac{\alpha}{\sqrt{\hbar^2-\theta^2}}
\Big[\lambda\big|(\tilde{p}_i-t\varepsilon_{ij}\tilde{x}_j)\phi\big)
-\sqrt{\hbar^2+\lambda^2-\theta^2}
\big|\phi(\tilde{x}_i+t\varepsilon_{ij}\tilde{p}_j)\big)\Big],\nonumber\\
&&\tilde{Q}_i|\phi)
=\frac{\alpha}{\sqrt{\hbar^2-\theta^2}}
\Big[\lambda\big|(\tilde{x}_i+t\varepsilon_{ij}\tilde{p}_j)\phi\big)
+\sqrt{\hbar^2+\lambda^2-\theta^2}
\big|\phi(\tilde{p}_i-t\varepsilon_{ij}\tilde{x}_j)\big)\Big],\qquad
\end{eqnarray}
where
\begin{equation}
\alpha=\frac{\sqrt{\hbar+\sqrt{\hbar^2-\theta^2}}}{\sqrt{2\hbar}}.
\end{equation}

The Dirac operator $\mathcal{D}$ for the generalized noncommutative phase space (\ref{ncps}) can be written as \cite{Gayral},
\begin{equation}\label{do0}
\mathcal{D}
=\frac{1}{\lambda}\gamma^1\tilde{Y}_1-\frac{1}{\lambda}\gamma^2\tilde{Q}_1
+\frac{1}{\lambda}\gamma^3\tilde{Y}_2-\frac{1}{\lambda}\gamma^4\tilde{Q}_2,
\end{equation}
where $\gamma^k$'s are the Euclidean Dirac matrices satisfying
$\gamma^k\gamma^l+\gamma^l\gamma^k=2\delta_{kl}\mathds{I}_4$, for example,
\begin{eqnarray}
&&\gamma^1=\left(
  \begin{array}{cccc}
    0 & 0 & 0 & \mathrm{i} \\
    0 & 0 & \mathrm{i} & 0 \\
    0 & -\mathrm{i} & 0 & 0 \\
    -\mathrm{i} & 0 & 0 & 0 \\
  \end{array}\right),
\qquad
\gamma^2=\left(
  \begin{array}{cccc}
    0 & 0 & 0 &1 \\
    0 & 0 & -1 & 0 \\
    0 & -1 & 0 & 0 \\
    1 & 0 & 0 & 0 \\
  \end{array}\right),
\nonumber\\
&&\gamma^3=\left(
  \begin{array}{cccc}
    0 & 0 & \mathrm{i} & 0 \\
    0 & 0 & 0 & -\mathrm{i} \\
    -\mathrm{i} & 0 & 0 & 0 \\
    0 & \mathrm{i} & 0 & 0 \\
  \end{array}\right),
\qquad
\gamma^4=\left(
  \begin{array}{cccc}
    0 & 0 & 1 & 0 \\
    0 & 0 & 0 & 1 \\
    1 & 0 & 0 & 0 \\
    0 & 1 & 0 & 0 \\
  \end{array}\right).
\end{eqnarray}
So the Dirac operator (\ref{do0}) can be expressed as,
\begin{equation}
\mathcal{D}
=\frac{1}{\lambda}\left(
  \begin{array}{cccc}
    0 & ~0 & \mathrm{i}\tilde{Y}_2-\tilde{Q}_2 & ~\mathrm{i}\tilde{Y}_1-\tilde{Q}_1 \\
    0 & ~0 & \mathrm{i}\tilde{Y}_1+\tilde{Q}_1 & ~-\mathrm{i}\tilde{Y}_2-\tilde{Q}_2 \\
    -\mathrm{i}\tilde{Y}_2-\tilde{Q}_2 & ~-\mathrm{i}\tilde{Y}_1+\tilde{Q}_1 & 0 & ~0 \\
    -\mathrm{i}\tilde{Y}_1-\tilde{Q}_1 & ~\mathrm{i}\tilde{Y}_2-\tilde{Q}_2 & 0 & ~0 \\
  \end{array}\right).
\end{equation}
After some straightforward calculations, one can obtain the commutator $[\mathcal{D},\pi(e)]$ acting on an element $\Phi\in \mathcal{Q}\otimes \mathbb{C}^4$ as
\begin{eqnarray}
[\mathcal{D},\pi(e)]\Phi
&=&[\mathcal{D},\pi(e)]
\left(
  \begin{array}{c}
   |\phi_1) \\
   |\phi_2) \\
   |\phi_3) \\
   |\phi_4) \\
  \end{array}
\right)\nonumber\\
&=&\beta\left(
  \begin{array}{cccc}
    \!0 & 0 &\![-\hat{a}_2^{\dag}{+}\mathrm{i}t\hat{a}_1\!{}^{\dag},e] & [-\hat{a}_1^{\dag}{-}\mathrm{i}t\hat{a}_2\!{}^{\dag},e] \\
    \!0 & 0 &\![\hat{a}_1{-}\mathrm{i}t\hat{a}_2,e] & [-\hat{a}_2{-}\mathrm{i}t\hat{a}_1,e] \\{}
    \![-\hat{a}_2{-}\mathrm{i}t\hat{a}_1,e] & [\hat{a}_1^{\dag}{+}\mathrm{i}t\hat{a}_2\!{}^{\dag},e] & \!0 & 0 \\{}
    \![-\hat{a}_1{+}\mathrm{i}t\hat{a}_2,e] & [-\hat{a}_2^{\dag}{+}\mathrm{i}t\hat{a}_1\!{}^{\dag},e] & \!0 & 0 \\
  \end{array}\!\right)
\!\!\left(
  \begin{array}{c}
   \!|\phi_1) \\
   \!|\phi_2) \\
   \!|\phi_3) \\
   \!|\phi_4) \\
  \end{array}
\!\right)\nonumber\\
&=&\beta\left(
  \begin{array}{cccc}
    0 & 0 & [-\hat{A}_2^\dag,e] & [-\hat{A}_1^\dag,e] \\
    0 & 0 &  [\hat{A}_1,e] & [-\hat{A}_2,e] \\{}
    [-\hat{A}_2,e] & [\hat{A}_1^\dag,e] & 0 & 0 \\{}
    [-\hat{A}_1,e] & [-\hat{A}_2^\dag,e] & 0 & 0 \\
  \end{array}\right)
\!\!\left(
  \begin{array}{c}
   \!|\phi_1) \\
   \!|\phi_2) \\
   \!|\phi_3) \\
   \!|\phi_4) \\
  \end{array}
\!\right).
\end{eqnarray}
So regarding $\Phi$ as a test function, one can identify the Dirac operator $\mathcal{D}$ as
\begin{equation}
\mathcal{D}=\beta\left(
  \begin{array}{cccc}
    0 & 0 & -\hat{A}_2^\dag & -\hat{A}_1^\dag \\
    0 & 0 &  \hat{A}_1 & -\hat{A}_2 \\
    -\hat{A}_2 & \hat{A}_1^\dag & 0 & 0 \\
    -\hat{A}_1 & -\hat{A}_2^\dag & 0 & 0 \\
  \end{array}\right).
\end{equation}

\section{Optimal elements of Connes distance between Fock states $|m,n\rangle$ and $|m+k,n+l\rangle$}\label{apb}
Using the expression (\ref{eo0}) and constraint relations (\ref{cr}), one can find some optimal elements corresponding to the Connes distance between Fock states $|m,n\rangle$ and $|m+k,n+l\rangle$ in the diagonal elements.

First, let us consider the distance between the states $|m+1,n+1\rangle$ and $|m,n\rangle$,
\begin{eqnarray}
\lefteqn{d(\omega_{m+1,n+1},\omega_{m,n})
=\sup_{e\in B}|\mathrm{tr}_{\mathcal{F}}(\rho_{m+1,n+1} e)-\mathrm{tr}_{\mathcal{F}}(\rho_{m,n} e)|}\nonumber\\
&=&\sup_{e\in B}|c_{m+1,n+1}-c_{m,n}|
=\sup_{e\in B}|E_{m+1,n+1}+F_{m,n+1}|.
\end{eqnarray}
From (\ref{cr}), there is
\begin{eqnarray}\label{ghgh}
&&(1+t^2)(G_{m+1,n}+H_{m,n+1})\leqslant\frac{1}{\beta^2},\qquad
(1+t^2)(G_{m+1,n+1}+H_{m+1,n+1})\leqslant\frac{1}{\beta^2},\nonumber\\
&& G_{m+1,n+1}+H_{m,n+1}\leqslant\frac{1}{\beta^2},\qquad G_{m+1,n}+H_{m+1,n+1}\leqslant\frac{1}{\beta^2}.
\end{eqnarray}

In order to attain the supremum of $|E_{m+1,n+1}+F_{m,n+1}|$, one must let $G_{m+1,n+1}+H_{m,n+1}$ as large as possible.
But if
\begin{equation}\label{gh1}
G_{m+1,n+1}+H_{m,n+1}>\frac{1}{(1+t^2)\beta^2},
\end{equation}
then compare with (\ref{ghgh}), there must be
\begin{equation}
G_{m+1,n}<G_{m+1,n+1},\qquad H_{m+1,n+1}<H_{m,n+1}.
\end{equation}
Without loss of generality, we can assume that $E_{m+1,n+1},F_{m,n+1}\geqslant0$. Therefore there are
\begin{equation}\label{eeff}
E_{m+1,n}< E_{m+1,n+1},\qquad
F_{m+1,n+1}< F_{m,n+1}.
\end{equation}
But
\begin{equation}
E_{m+1,n}+F_{m+1,n+1}=c_{m+1,n+1}-c_{m,n}=E_{m+1,n+1}+F_{m,n+1},
\end{equation}
this is in contradiction with the relations (\ref{eeff}). This means that the inequality (\ref{gh1}) is not true.

So there is
\begin{equation}
G_{m+1,n+1}+H_{m,n+1}\leqslant\frac{1}{(1+t^2)\beta^2},
\end{equation}
In order to attain the supremum of $|E_{m+1,n+1}+F_{m,n+1}|$, there must be
\begin{equation}
 G_{m+1,n+1}+H_{m,n+1}=\frac{1}{(1+t^2)\beta^2},
\end{equation}
and
\begin{equation}
G_{m+1,n}=G_{m+1,n+1},\qquad H_{m+1,n+1}=H_{m,n+1},
\end{equation}
or
\begin{equation}
E_{m+1,n}=E_{m+1,n+1},\qquad
F_{m+1,n+1}=F_{m,n+1}.
\end{equation}

So the Connes spectral distance between the states $|m+1,n+1\rangle$ and $|m,n\rangle$ is
\begin{eqnarray}
\lefteqn{d(\omega_{m+1,n+1},\omega_{m,n})
=\sup_{e\in B}|E_{m+1,n+1}+F_{m,n+1}|}\nonumber\\
&=&\sup_{e\in B}\left|\frac{1}{\sqrt{m+1}}E_{m+1,n+1}\sqrt{m+1}
+\frac{1}{\sqrt{n+1}}F_{m,n+1}\sqrt{n+1}\right|\nonumber\\
&\leqslant&\sup_{e\in B}\sqrt{\frac{1}{m+1}+\frac{1}{n+1}}
\sqrt{(m+1)E_{m+1,n+1}^2+(n+1)F_{m,n+1}^2}\nonumber\\
&=&\sup_{e\in B}\sqrt{\frac{1}{m+1}+\frac{1}{n+1}}
\sqrt{G_{m+1,n+1}+H_{m,n+1}}\nonumber\\
&=&\sqrt{\frac{1}{m+1}+\frac{1}{n+1}}\sqrt{\frac{1}{(1+t^2)\beta^2}}.
\end{eqnarray}
In the above inequality, we have used the Cauchy-Schwartz inequality,
\begin{equation}\label{cs}
(x_1y_1+x_2y_2)^2\leqslant(x_1^2+x_2^2)(y_1^2+y_2^2),
\end{equation}
where the equality holds if $x_1 y_2=x_2 y_1$.

So one can set
\begin{equation}
(m+1)E_{m+1,n+1}=(n+1)F_{m,n+1},
\end{equation}
and there is
\begin{equation}
d(\omega_{m+1,n+1},\omega_{m,n})
=\frac{1}{\beta\sqrt{1+t^2}}\sqrt{\frac{1}{m+1}+\frac{1}{n+1}}.
\end{equation}

Furthermore, let us study the optimal elements corresponding to Connes spectral distance between two Fock states $|m,n\rangle$ and $|m+k,n+l\rangle$. For simplicity, we assume $k,l>1$. The corresponding Connes spectral distance is
\begin{eqnarray}
\lefteqn{d(\omega_{m+k,n+l},\omega_{m,n})
=\sup_{e\in B}|\mathrm{tr}_{\mathcal{F}}(\rho_{m+k,n+l}\,e)-\mathrm{tr}_{\mathcal{F}}(\rho_{m,n}\,e)|}\nonumber\\
&=&\sup_{e\in B}|c_{m+k,n+l}-c_{m,n}|
=\sup_{e\in B}|c_{m+k,n+l}-c_{m+k,n}+c_{m+k,n}-c_{m,n}|\nonumber\\
&=&\sup_{e\in B}\left|\sum_{i=1}^{l}(c_{m+k,n+i}-c_{m+k,n+i-1})
+\sum_{j=1}^{k}(c_{m+j,n}-c_{m+j-1,n})\right|\nonumber\\
&=&\sup_{e\in B}\left|\sum_{i=1}^{l}F_{m+k,n+i}+\sum_{j=1}^{k}E_{m+j,n}\right|.
\end{eqnarray}
From (\ref{cr}), in order to attain the supremum of $|c_{m+k,n+l}-c_{m,n}|$,
there must be
\begin{eqnarray}
&& (1+t^2)(G_{m+2,n+1}+H_{m+1,n+2})=\frac{1}{\beta^2},\quad
(1+t^2)(G_{m+1,n+1}+H_{m+1,n+1})=\frac{1}{\beta^2},\nonumber\\
&& (G_{m+2,n+1}+H_{m+1,n+1})+t^2(G_{m+1,n+1}+H_{m+1,n+2})=\frac{1}{\beta^2},\nonumber\\
&& (G_{m+1,n+1}+H_{m+1,n+2})+t^2(G_{m+2,n+1}+H_{m+1,n+1})=\frac{1}{\beta^2}.
\end{eqnarray}
So there is
\begin{equation}
G_{m+2,n+1}=G_{m+1,n+1},\qquad
H_{m+1,n+1}=H_{m+1,n+2}.
\end{equation}
Similarly, one can obtain
\begin{eqnarray}
&&G_{m+i,n+j}=G_{m+1,n},
\qquad 1\leqslant i\leqslant k,0\leqslant j\leqslant l,\nonumber\\
&&H_{m+i,n+j}=H_{m,n+1},
\qquad 0\leqslant i\leqslant k,1\leqslant j\leqslant l,
\end{eqnarray}
or
\begin{equation}
E_{m+i,n+j}=\sqrt{\frac{m+1}{m+i}}E_{m+1,n},
\qquad
F_{m+i,n+j}=\sqrt{\frac{n+1}{n+j}}F_{m,n+1}.
\end{equation}
We also have
\begin{equation}
(1+t^2)(G_{m+1,n}+H_{m,n+1})=\frac{1}{\beta^2},
\end{equation}
or
\begin{equation}\label{me1}
(m+1)E_{m+1,n}^2+(n+1)F_{m,n+1}^2=\frac{1}{(1+t^2)\beta^2}.
\end{equation}

Therefore the Connes spectral distance between the Fock states $|m,n\rangle$ and $|m+k,n+l\rangle$ is
\begin{eqnarray}
\lefteqn{d(\omega_{m+k,n+l},\omega_{m,n})
=\sup_{e\in B}\left|\sum_{i=1}^{k}E_{m+i,n}+\sum_{j=1}^{l}F_{m+k,n+j}\right|}\nonumber\\
&=&\sup_{e\in B}\left|\sqrt{m+1}E_{m+1,n}\sum_{i=1}^{k}\frac{1}{\sqrt{m+i}}
+\sqrt{n+1}F_{m,n+1}\sum_{j=1}^{l}\frac{1}{\sqrt{n+j}}\right|\nonumber\\
&=&\sup_{e\in B}\left|\sqrt{m+1}E_{m+1,n}\zeta_{m;m+k}
+\sqrt{n+1}F_{m,n+1}\zeta_{n;n+l}\right|\nonumber\\
&\leqslant&\sup_{e\in B}
\sqrt{(m+1)E_{m+1,n}^2+(n+1)F_{m,n+1}^2}
\sqrt{\zeta_{m;m+k}^2+\zeta_{n;n+l}^2}\nonumber\\
&=&\frac{1}{\beta\sqrt{1+t^2}}
\sqrt{\zeta_{m;m+k}^2+\zeta_{n;n+l}^2}.
\end{eqnarray}
Here we have used the Cauchy-Schwartz inequality (\ref{cs}) in the above inequality, and the equality holds if
\begin{equation}\label{ef}
\sqrt{m+1}E_{m+1,n}\zeta_{n;n+l}=\sqrt{n+1}F_{m,n+1}\zeta_{m;m+k}.
\end{equation}
From (\ref{me1}) and (\ref{ef}), one can derive
\begin{equation}
E_{m+1,n}=\frac{1}{\beta\sqrt{1+t^2}}\frac{1}{\sqrt{m+1}}
\frac{\zeta_{m;m+k}}{\sqrt{\zeta_{m;m+k}^2
+\zeta_{n;n+l}^2}}=c_0\frac{1}{\sqrt{m+1}},
\end{equation}
\begin{equation}
F_{m,n+1}=\frac{1}{\beta\sqrt{1+t^2}}\frac{1}{\sqrt{n+1}}
\frac{\zeta_{n;n+l}}{\sqrt{\zeta_{m;m+k}^2
+\zeta_{n;n+l}^2}}=d_0\frac{1}{\sqrt{n+1}},
\end{equation}
where
\begin{equation}
c_0=\frac{1}{\beta\sqrt{1+t^2}}
\frac{\zeta_{m;m+k}}{\sqrt{\zeta_{m;m+k}^2
+\zeta_{n;n+l}^2}},\qquad
d_0=\frac{1}{\beta\sqrt{1+t^2}}
\frac{\zeta_{n;n+l}}{\sqrt{\zeta_{m;m+k}^2
+\zeta_{n;n+l}^2}}.
\end{equation}

So in order to attain the supremum of $|c_{m+k,n+l}-c_{m,n}|$, for example, one can choose
\begin{eqnarray}
c_{p,q}&=&c_0\sum_{i=1}^{p}\frac{1}{\sqrt{i}}
+d_0\sum_{j=1}^{q}\frac{1}{\sqrt{j}}
=c_0\zeta_{0;p}+d_0\zeta_{0;q}\nonumber\\
&=&\frac{1}{\beta\sqrt{1+t^2}}
\frac{\zeta_{0;p}\zeta_{m;m+k}+\zeta_{0;q}\zeta_{n;n+l}}{\sqrt{\zeta_{m;m+k}^2
+\zeta_{n;n+l}^2}},
\end{eqnarray}
where $p=m,m+1,...,m+k$ and $q=n,n+1,...,n+l$\,.
Obviously, there is
\begin{equation}
E_{i+1,j}=c_{i+1,j}-c_{i,j}=\frac{1}{\sqrt{i+1}}c_0,
\qquad
F_{i,j+1}=c_{i,j+1}-c_{i,j}=\frac{1}{\sqrt{j+1}}d_0.
\end{equation}
The corresponding optimal element $e_o$ is
\begin{eqnarray}\label{eo1}
e_o&=&\sum_{p=m}^{m+k}\sum_{q=n}^{n+l}
c_{p,q}|p,q\rangle\langle p,q|\nonumber\\
&=&\frac{1}{\beta\sqrt{1+t^2}}\sum_{p=m}^{m+k}\sum_{q=n}^{n+l}
\frac{\zeta_{0;p}\zeta_{m;m+k}+\zeta_{0;q}\zeta_{n;n+l}}{\sqrt{\zeta_{m;m+k}^2
+\zeta_{n;n+l}^2}}|p,q\rangle\langle p,q|,
\end{eqnarray}
and the spectral distance between the Fock states $|m+k,n+l\rangle$ and $|m,n\rangle$ is
\begin{eqnarray}\label{d2}
d(\omega_{m+k,n+l},\omega_{m,n})
&=&|\mathrm{tr}_{\mathcal{F}}(\rho_{m+k,n+l}\, e_o)-\mathrm{tr}_{\mathcal{F}}(\rho_{m,n}\, e_o)|\nonumber\\
&=&|\langle m+k,n+l|e_o|m+k,n+l\rangle-\langle m,n|e_o|m,n\rangle|\nonumber\\
&=&c_0\zeta_{m;m+k}+d_0\zeta_{n;n+l}\nonumber\\
&=&\frac{1}{\beta\sqrt{1+t^2}}
\sqrt{\zeta_{m;m+k}^2+\zeta_{n;n+l}^2}.
\end{eqnarray}
It is easy to verify that, these results (\ref{eo1}) and (\ref{d2}) are also true for $k\leqslant0$ and/or $l\leqslant0$.


\begin{thebibliography}{100}

\bibitem{Snyder} H. S. Snyder, ``Quantized space-time.'' {\it Phys. Rev.} {\bf 71} 38 (1947).

\bibitem{Connes} A. Connes, {\it Noncommutative Geometry} (Academic Press, New York, 1994).

\bibitem{Seiberg} N. Seiberg, E. Witten, ``String theory and noncommutative geometry.'' {\it J. High Energy Phys.} {\bf 09} 032 (1999).

\bibitem{Douglas} M. R. Douglas, N. A. Nekrasov, ``Noncommutative field theory.'' {\it Rev. Mod. Phys.} {\bf 73} 977 (2001).

\bibitem{Gayral} V. Gayral, J. M. Gracia-Bond\'{\i}a, B. Iochum, T. Sch\"{u}cker, J. C. V\'{a}rilly, ``Moyal planes are spectral triple.'' {\it Commun. Math. Phys.} {\bf 246} 569–623 (2004).

\bibitem{Chaichian} M. Chaichian, P. Pre\v{s}najder, A. Tureanu, ``New concept of relativistic invariance in noncommutative space-time: twisted Poincar\'{e} symmetry and its implications.'' {\it Phys. Rev. Lett.} {\bf 94} 151602 (2005).

\bibitem{Ettefaghi} M. M. Ettefaghi, M. Haghighat, ``Massive neutrino in noncommutative space-time.'' {\it Phys. Rev. D} {\bf 77} 056009 (2008).

\bibitem{lhc} B. S. Lin, T. H. Heng, W. Chen, ``Quantum field theory with a minimal length induced from noncommutative space.'' {\it Commun. Theor. Phys.} {\bf 61} 605 (2014).

\bibitem{Calmet} X. Calmet, C. Fritz, ``Inflation on a non-commutative space-time.'' {\it Phys. Lett. B} {\bf 747} 406 (2015).

\bibitem{Couch} J. Couch, S. Eccles, W. Fischler, M.-L. Xiao, ``Holographic complexity and noncommutative gauge theory.'' {\it J. High Energy Phys.} {\bf 03} 108 (2018).

\bibitem{Muhuri} A. Muhuri, D. Sinha, S. Ghosh, ``Entanglement induced by noncommutativity: anisotropic harmonic oscillator in noncommutative space.'' {\it Eur. Phys. J. Plus} {\bf 136} 35 (2021).

\bibitem{Doplicher} S. Doplicher, K. Fredenhagen, J. E. Roberts, ``Spacetime quantization induced by classical gravity.'' {\it Phys. Lett. B} {\bf 331} 39 (1994).

\bibitem{Zupnik} B. M. Zupnik, ``Reality in noncommutative gravity.'' {\it Class. Quantum Grav.} {\bf 24} 15 (2007).

\bibitem{Polychronakos} A. P. Polychronakos, ``Quantum Hall states on the cylinder as unitary matrix Chern-Simons theory.'' {\it J. High Energy Phys.} {\bf 06} 070 (2001).

\bibitem{Duval} C. Duval, P. A. Horv\'{a}thy, ``The exotic Galilei group and the `Peierls substitution'.'' {\it Phys. Lett. B} {\bf 479} 284 (2000).

\bibitem{Nair} V. P. Nair, A. P. Polychronakos, ``Quantum mechanics on the noncommutative plane and sphere.'' {\it Phys. Lett. B} {\bf 505} 267 (2001).

\bibitem{Banerjee} R. Banerjee, ``A novel approach to noncommutativity in planar quantum mechanics.'' {\it Mod. Phys. Lett. A} {\bf 17} 631 (2002).

\bibitem{Zhang} J. Z. Zhang, ``Fractional angular momentum in non-commutative spaces.'' {\it Phys. Lett. B} {\bf 584} 204 (2004).

\bibitem{Li} K. Li, J. H. Wang, C. Y. Chen, ``Representation of noncommutative phase space.'' {\it Mod. Phys. Lett. A} {\bf 20} 2165 (2005).

\bibitem{Bastos} C. Bastos, O. Bertolami, N. C. Dias, J. N. Prata, ``Phase-space noncommutative quantum cosmology.'' {\it Phys. Rev. D} {\bf 78} 023516 (2008).

\bibitem{HMS} P. A. Horv\'{a}thy, L. Martina, P. C. Stichel, ``Exotic Galilean symmetry and non-commutative mechanics.'' {\it SIGMA} {\bf 6} 060 (2010).

\bibitem{Jing} S. C. Jing, B. S. Lin, ``A new kind of representations on noncommutative phase space.'' {\it Phys. Lett. A} {\bf 372} 7109 (2008).

\bibitem{Lin} B. S. Lin, T. H. Heng, ``Energy spectra of the harmonic oscillator in a generalized noncommutative phase space of arbitrary dimension.'' {\it Chin. Phys. Lett.} {\bf 28} 070303 (2011).

\bibitem{Gnatenko} Kh. P. Gnatenko, O. V. Shyiko, ``Effect of noncommutativity on the spectrum of free particle and harmonic oscillator in rotationally invariant noncommutative phase space.'' {\it Mod. Phys. Lett. A} {\bf 33} 1850091 (2018).

\bibitem{Connes1} A. Connes, ``Compact metric spaces, Fredholm modules and hyperfiniteness.'' {\it Ergod. Theory Dyn. Syst.} {\bf 9} 207-220 (1989).

\bibitem{Bimonte} G. Bimonte, F. Lizzi, G. Sparano, ``Distances on a lattice from non-commutative geometry.'' {\it Phys. Lett. B} {\bf 341} 139-146 (1994).

\bibitem{Cagnache} E. Cagnache, F. D'Andrea, P. Martinetti, J.-C. Wallet, ``The spectral distance on the Moyal plane.'' {\it J. Geom. Phys.} {\bf 61} 1881-1897 (2011).

\bibitem{Martinetti} P. Martinetti, L. Tomassini, ``Noncommutative geometry of the Moyal plane: translation isometries, Connes’ distance on coherent states, Pythagoras equality.'' {\it Commun. Math. Phys.} {\bf 323} 107–141 (2013).

\bibitem{DAndrea0} F. D’Andrea, F. Lizzi, J.C. V\'{a}rilly, ``Metric properties of the fuzzy sphere.'' {\it Lett. Math. Phys.} {\bf 103} 183–205 (2013).

\bibitem{DAndrea} F. D’Andrea, P. Martinetti, ``On Pythagoras theorem for products of spectral triples.'' {\it Lett. Math. Phys.} {\bf 103} 469–492 (2013).

\bibitem{Franco} N. Franco, J.-C. Wallet, ``Metrics and causality on Moyal planes.'' {\it Contemp. Math.} {\bf 676} 147-173 (2016).

\bibitem{Scholtz} F. G. Scholtz, B. Chakraborty, ``Spectral triplets, statistical mechanics and emergent geometry in non-commutative quantum mechanics.'' {\it J. Phys. A: Math. Theor.} {\bf 46} 085204 (2013).

\bibitem{Chaoba} Y. Chaoba Devi, S. Prajapat, A. K. Mukhopadhyay, B. Chakraborty, F. G. Scholtz, ``Connes distance function on fuzzy sphere and the connection between geometry and statistics.'' {\it J. Math. Phys.} {\bf 56} 041707 (2015).

\bibitem{Revisiting} Y. Chaoba Devi, K. Kumar, B. Chakraborty, F. G. Scholtz, ``Revisiting Connes’ finite spectral distance on noncommutative spaces: Moyal plane and fuzzy sphere.'' {\it Int. J. Geom. Methods Mod. Phys.} {\bf 15} 1850204 (2018).

\bibitem{Kumar} K. Kumar, B. Chakraborty, ``Spectral distances on the doubled Moyal plane using Dirac eigenspinors.'' {\it Phys. Rev. D} {\bf 97} 086019 (2018).

\bibitem{Barrett} J. W Barrett, P. Druce, L. Glaser, ``Spectral estimators for finite non-commutative geometries.'' {\it J. Phys. A: Math. Theor.} {\bf 52} 275203 (2019).

\bibitem{Chakraborty} A. Chakraborty, B. Chakraborty, ``Spectral distance on Lorentzian Moyal plane.'' {\it Int. J. Geom. Methods Mod. Phys.} {\bf 17} 2050089 (2020).

\bibitem{Lin1} B. S. Lin, J. Xu, T. H. Heng, ``Induced entanglement entropy of harmonic oscillators in non-commutative phase space.'' {\it Mod. Phys. Lett. A} {\bf 34} 1950269 (2019).

\bibitem{squeezed} B. S. Lin, S. C. Jing, ``Deformed squeezed states in noncommutative phase space.'' {\it Phys. Lett. A} {\bf 372} 4880 (2008).

\bibitem{Formulation} F. G. Scholtz, L. Gouba, A. Hafver, C. M. Rohwer, ``Formulation, interpretation and application of non-commutative quantum mechanics.'' {\it J. Phys. A: Math. Theor.} {\bf 42} 175303 (2009).

\bibitem{Iochum} B. Iochum, T. Krajewski, P. Martinetti, ``Distances in finite spaces from noncommutative geometry.'' {\it J. Geom. Phys.} {\bf 37} 100–125 (2001).



\end{thebibliography}
\end{document}